\input psfig.tex

\def\th{{\thinspace}}
\def\Mo{{$M_\odot $}}
\def\Lo{{$L_\odot $}}
\def\lgt{ \raise4pt \hbox{$<$}\kern-9pt\lower1.5pt \hbox{$>$}}
\def\glt{\raise4pt \hbox{$<$}\kern-9pt\lower1.5pt\hbox{$>$}}
\def\approxgt{\raise3pt\hbox{${\scriptstyle>}$}
           \kern-6pt\lower1.1pt\hbox{${\scriptstyle\sim}$}}
\def\approxlt{\raise3pt\hbox{${\scriptstyle<}$}
           \kern-6pt\lower1.1pt\hbox{${\scriptstyle\sim}$}}
\def\ni{\noindent}
\def\dotd{\hbox{$.\!\!^{\rm d}$}}

 \documentstyle[11pt,conf_iap]{article}
 \begin{document}
 \heading{CLASSICAL CEPHEIDS -- A REVIEW}

 \vskip 10pt

 \centerline{\psfig{figure=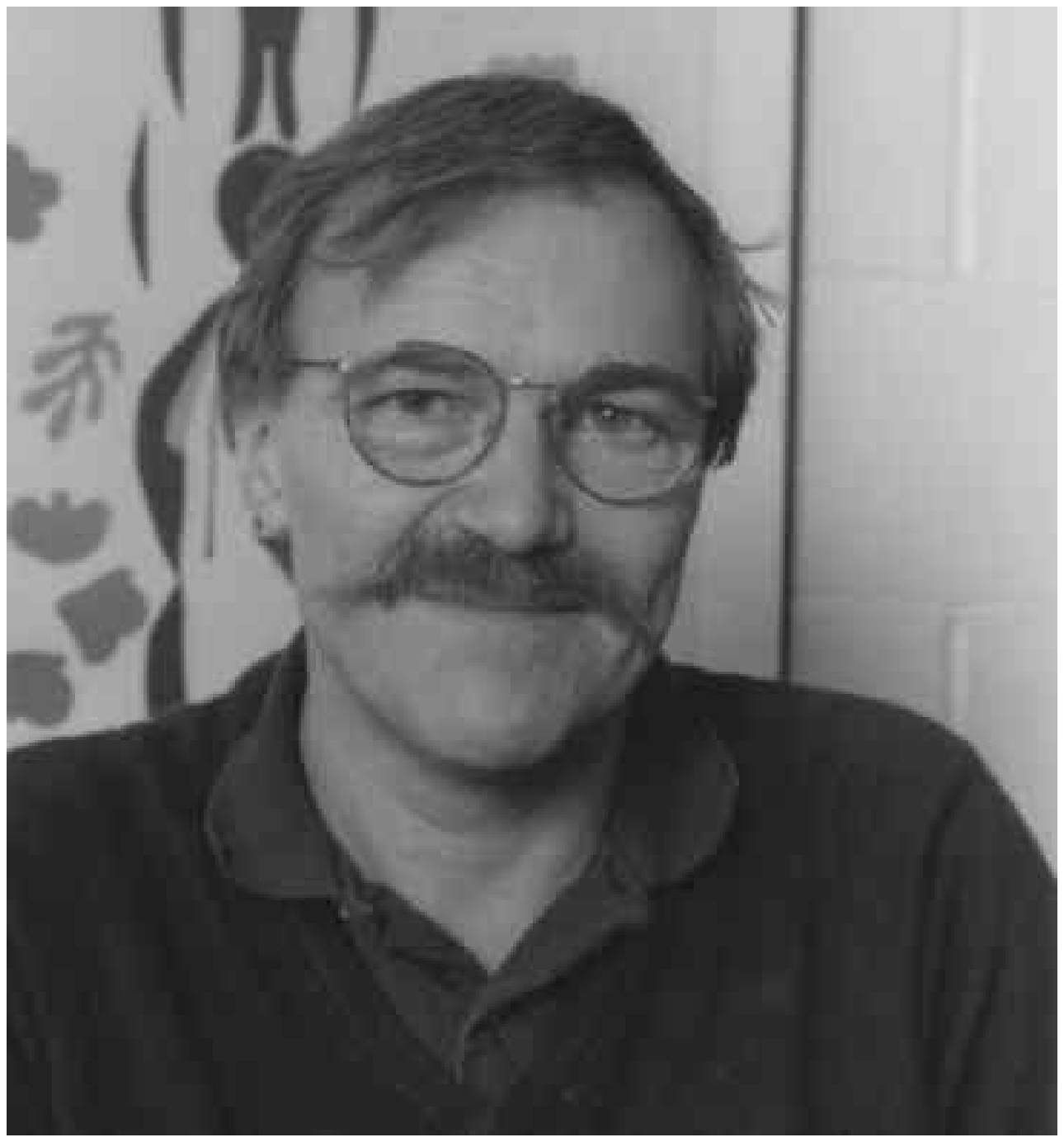,height=4.1truecm}}
 \author{J. Robert BUCHLER}
       {$^{1}$ Physics Department, University of Florida, 
Gainesville FL32611, USA}

 \bigskip

 \begin{abstract}{\baselineskip 0.4cm
 We briefly review the evolutionary status
of the classical Cepheid variables, their structure, and the properties of the
linear modes.  Then we discuss the current status of the nonlinear
hydrodynamical modelling, including modern adaptive mesh techniques, and the
sensitivity to opacity.  We stress the importance of resonances among the
normal modes of pulsation because they can be used to impose severe constraints
on the models and on the input physics.

 We show that recent hydrodynamics computations provide an explanation for the
hitherto unexplained dip in the histogram of Galactic Cepheids around
8--10$^d$.

 The EROS and MACHO observations of classical Cepheids in the Large and Small
Magellanic Clouds have opened up a new exciting perspective on these variable
stars.  This new observational basis is very important, not only because of the
sheer number of new Cepheids, but especially because the composition of the MCs
is known to differ from our Galaxy's, thus extending the range of astrophysical
parameters.  These new data have been found not to be fully compatible with our
current Cepheid modelling and are forcing a review of our assumptions and input
physics.
  } \end{abstract}

\section{INTRODUCTION}

This Colloquium to within a month marks the fourth centennial of variable star
Astronomy.  Indeed, in 1596, at the time of Tycho Brahe, and a dozen years
before the development of the telescope, a priest in Friesland, David
Fabricius, recorded that a star in the constellation of the Whale that was
quite bright in August had disappeared from sight by the end of the year, only
to reappear with full brightness the following year.  Because of its unusual
nature this star was called Stella Mirabilis (the 'remarkable star') or Mira
for short, and it is now a prototype of a whole class of variable stars.
Almost another 200 years elapsed before a young Englishman, John Goodricke,
discoved the variability of $\delta$ Cephei, which came to lend its name to the
class of variable stars which interest us here.  Cepheids are perhaps the best
known variable stars because of the relatively tight relation between period
and luminosity that Leavitt discovered in 1912, and which catapulted the
Cepheids into the prominent cosmological role of distance indicators.

\section{STRUCTURE AND LINEAR PROPERTIES}

Cepheids are stars that have evolved away from the Main Sequence.  The
evolutionary tracks for stars in the Cepheid mass range typically cross the
{\sl instability strip} three times ({\sl e.g.} \cite{Schaller}, \cite{BIT}).
In this instability strip the stars are vibrationally unstable and exhibit
pulsations.  Generally the first crossing is ignored as far as Cepheid models
are concerned.  The rationale is that, since the evolution is much more rapid
there than on the subsequent horizontal loops, there is a very low probability
of observing Cepheids on the first crossing.  However, the large number of
Cepheids that are being observed in the Magellanic Clouds make the possibility
of detection increasingly likely.  A search for such first crossers could yield
interesting new information about the structure of the Cepheid variables.

Evolutionary calculations completely ignore pulsation because the pulsation is
essentially a surface phenomenon - see below - that cannot have a large effect
on the evolution.  On the other hand, because the (nuclear) evolution time scale
is very much longer than the dynamical or thermal time scales the pulsation
studies can in turn ignore evolution.  As far as pulsation studies are
concerned, evolution calculations merely provide information about the location
of the loops and thus about the luminosities of the stars as a function of
mass, the so-called Mass--Luminosity ($M$--$L$) relation.

That the Cepheid pulsation is a surface phenomenon is illustrated on the left
of Fig.~1.  The top shows the pressure and temperature runs of a Cepheid model
($M$=6\Mo, $L$=4000\Lo, $Z$=0.01, $X$=0.70) in terms of the normalized radius.
The pins point to the ionization fronts, {\sl i.e.}  the points where the
relative mass fraction of the indicated ion is 50\%.  The very sharp drop in
temperature is caused by the hydrogen ionization front and is notorious for the
numerical headaches it has caused in the hydrodynamics.  

The bottom graph displays the modulus of the absolute radial displacement
eigenvector (scaled by the stellar radius) for the fundamental mode and the
first two overtones, having 0, 1 and 2 minima, respectively.  The bars on top
indicate the exterior mass fraction.  While the outer 60\% of the
radius partakes in the pulsation, it is less than 1\% of the mass that is
affected.

The right side of Fig.~1 shows the work integrands for the fundamental mode and
first overtone.  The driving which occurs in the partial ionization regions
(\cite{CoxGiuli}) is seen to be localized to the outer $\sim$0.01\% of the
mass, or 10\% of the radius.

\centerline{\psfig{figure=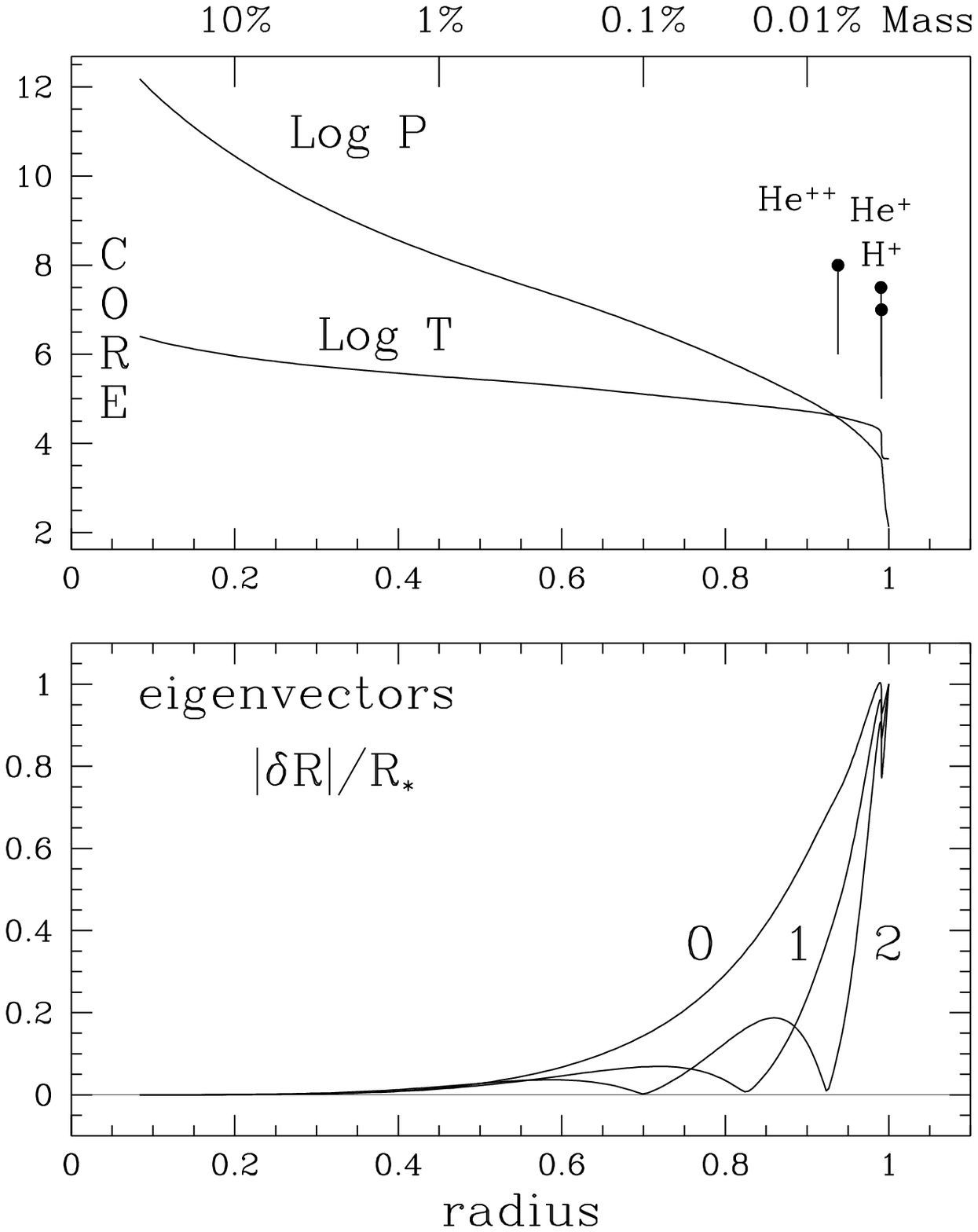,height=10cm}
\psfig{figure=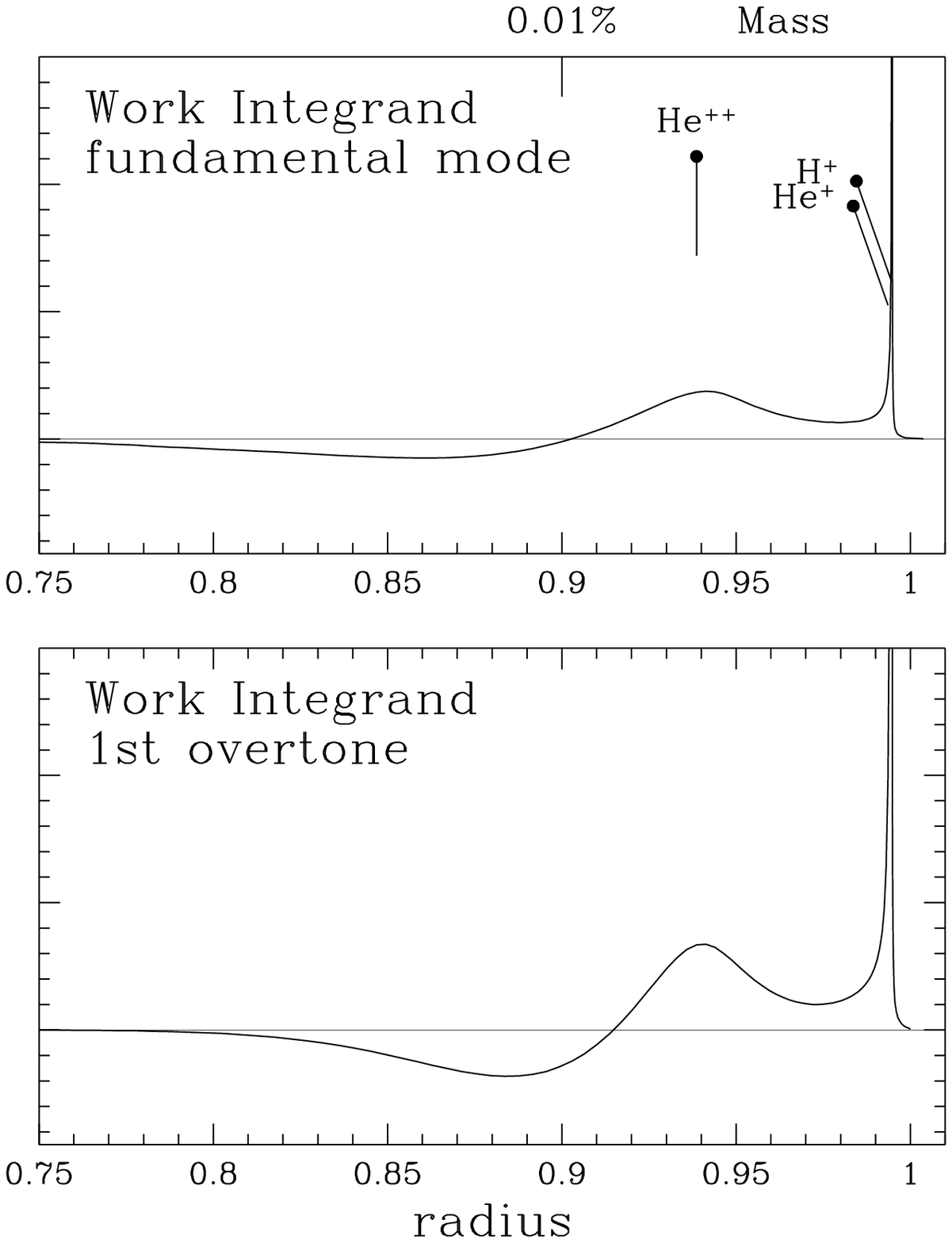,height=10cm}}

\ni{\footnotesize Fig.~1\ {\sl left}: Pressure and temperature profile,
eigenvectors {\sl vs} normalized radius, and {\sl vs} exterior mass (top
scale); {\sl right:} work integrands (cycle averaged $<p dV>$).}

  \vskip 10pt

\section {NONLINEAR PULSATIONS}

\subsection{Bump Progression and Fourier decomposition}

Resonances are known to play an important role in shaping the morphology of the
light- and radial velocity curves.  In 1926 Hertzsprung noted that in the
vicinity of a period of 10$^d$ the fundamental mode Cepheid light curves
exhibit a secondary maximum or shoulder that moves from descending to ascending
branch as the period increases.  This bump progression is even more striking
\cite{SimonMoffett} in the phases $\phi_{k1}=\phi_k-k\phi_1$ and amplitude
ratios $R_{k1}=A_k/A_1$ of the Fourier decomposition
 \begin{equation}
 f(t) = A_0 + \sum_{k=1,\ldots} A_k {\rm cos}(k \omega t
+\phi_k)
      =A_0 + A_1 \Bigg(\sum_{k=1,\ldots} R_{k1} {\rm cos}(k \omega t'
+\phi_{k1})\Bigg)
 \end{equation}
 ($t'=t-t_o$, where the epoch $t_o$ is such that $\phi_1=0$).  The advantage of
these {\sl relative} quantities $\phi_{k1}$ and $R_{k1}$ is that the
decomposition becomes independent of the origin of time and of the overall
pulsation amplitude.  On the basis of a comparison of linear models and
hydrodynamics Simon \& Schmidt \cite{SimonS} conjectured that the bump
progression is the result of a 2:1 resonance ($2\omega_0\approx \omega_2$)
between the fundamental mode of pulsation and the second overtone.

The presence of a resonance such as this puts a severe constraint on the
modelling, a crude sort of asteroseismology.  In fact, for a number of years
Cepheid modelling was plagued by a so-called 'mass discrepancy', which was the
inability of evolutionary $M$--$L$ relations to put the 2:1 resonance at the
observed period of 10$^d$.  Simon's \cite{Simonopac} bold suggestion that the
opacities had to be wrong initiated a large revision of the opacity tables
\cite{IR} \cite{SYMP}.  Calculations with these new opacities, supplemented
with the low temperature ones of \cite{Alex}, showed essential agreement
between evolution, linear pulsation theory and observations \cite{MBM}\cite{KS}
for the Galactic Cepheids.  This is one of those remarkable occasions of useful
positive feedback between astrophysics and physics.

\subsection {Numerical Hydrodynamics}

Numerical hydrodynamical modelling goes back to Christy in the 1960s.  These
early calculations reproduced the bulk properties of the Galactic Cepheids
reasonably well, except for the mass discrepancy.  With the new opacities, the
complete agreement between the observations and the hydrodynamical computations
is good, both for the light- and for the radial velocity curves \cite{MBM}.

While numerical hydrodynamics reproduce the bump and the Fourier coefficient
progression quite well, they provide no real explanation for the underlying
mechanism.  For this purpose the general amplitude equation formalism was
developed (for a review and references {\sl cf.} \cite{BuchlerFl}
\cite{BuchlerMito}).  These amplitude equations describe the nonlinear
interaction of the excited modes and give the behavior of the Fourier
coefficients as a function of period.

\subsection{Amplitude Equations}

This semi-analytical formalism takes advantage of the two small parameters with
which the Cepheids are endowed, namely (1) the ratio $\eta$ of growth rate to
frequency for the excited mode(s) and (2) the weak anharmonicity of the
pulsations (which allows an expansion in the pulsation amplitude).  Together
they allow one to find a beautifully simple description of the pulsation.

The formalism has been described in detail elsewhere \cite{BuchlerMito}.  Here,
as an illustration, we would like to just mention the simplest possible
situation, namely that of a single excited mode.  Let its linear frequency and
growth rate be $\omega$ and $\kappa$.  Furthermore, let $A$ denote the
pulsation amplitude.  The amplitude equation which governs the behavior of the
pulsation is
 \begin{equation}
 {dA\over dt} = \kappa A - Q A^3 + O(A^5), 
 \end{equation}
 where $Q$ is a nonlinear quantity that depends on the structure of the star.

If the star is linearly stable, $\kappa < 0$ and the long term solution of the
equation is $A(t)\rightarrow 0$.  In the case of linear instability, on the
other hand, the star approaches a constant pulsation amplitude given by
$A_{lc}=\sqrt{\kappa/Q}$, \th {\sl i.e}\th a limit cycle pulsation. (For
simplicity we have assumed that all the other modes are linearly stable.)
Finally, all stellar variables, such as radius, velocity, temperature,
luminosity, {\sl etc.}, are expressible in terms of $A_{lc}$,\th {\sl e.g.}
 \begin{equation}
 m_*(t)\th - <m> = A \th {\rm cos} \bar\omega t + \Bigg((*) A^2 + (*) A^2 {\rm
cos} 2\bar\omega t \Bigg) + O(A^3).
 \end{equation}
 Here $\bar\omega=\omega$ + a nonlinear correction.  The formalism generates
the successive terms of a Fourier expansion.  This is obviously a useful
property since such an expansion is the natural way to quantify the light curve
as we have seen in Eq.~1.

Of course, the formalism is more interesting and useful when several modes are
coupled, and in particular when they are coupled through a resonance
\cite{BuchlerFl} \cite{BuchlerMito}.  Thus, for example, amplitude equations
appropriate for the resonant coupling ($2\omega_0\approx \omega_2$) of the
fundamental mode with the second overtone provide an excellent description and
explanation of the light- and radial velocity curves of the bump Cepheids
throughout the whole resonance region.

\subsection{Effect of the Metallicity}

Because of the difference in metallicity between the Galaxy and the Magellanic
Clouds the behavior of the light curves as a function of $Z$ is of particular
interest.  In Fig.~2. we show some preliminary results of a survey of
fundamental Cepheid pulsations (Piciullo, Buchler, Koll\'ath \& Goupil, in
progress).  The top left figure displays the behavior of the magnitude Fourier
phases as a function of nonlinear period for three values of metallicity $Z$
for fixed $X$=0.7 \th (The lines are merely there to visually connect the
computed points).  The $M$--$L$ relations (with slope 1/3.56) for all three
sequences have been chosen to have their 2:1 resonance at 10$^d$, and the
sequences run parallel to the blue edge, offset by 100\th K.  It is noteworthy
that in the resonance region the metallicity has a strong effect on the phases
and on the amplitude ratios.  This sensitivity shows that care must be
exercised when the observational Fourier decomposition coefficients are used to
locate the resonance.

 \centerline{\psfig{figure=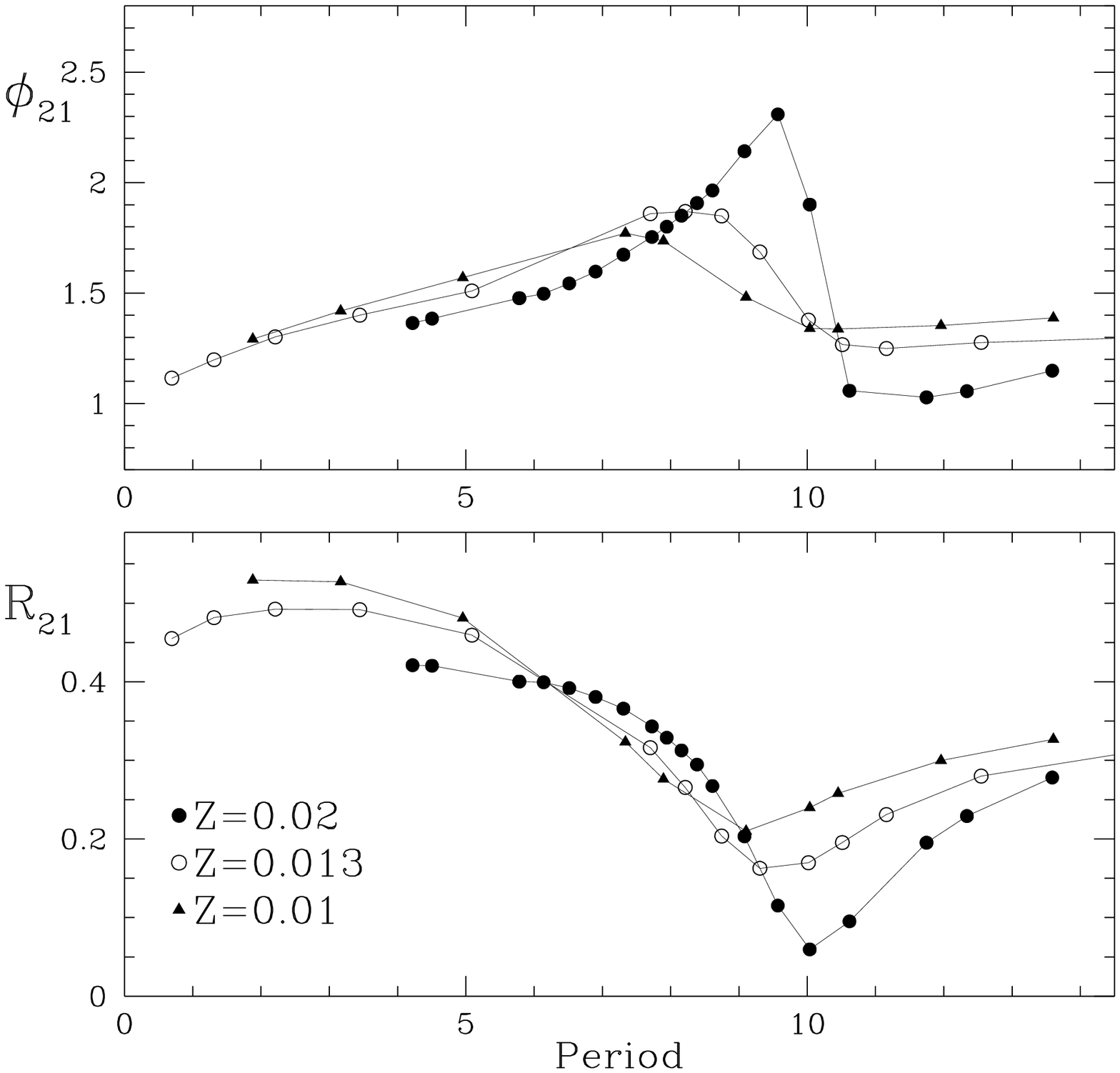,height=8cm}
\psfig{figure=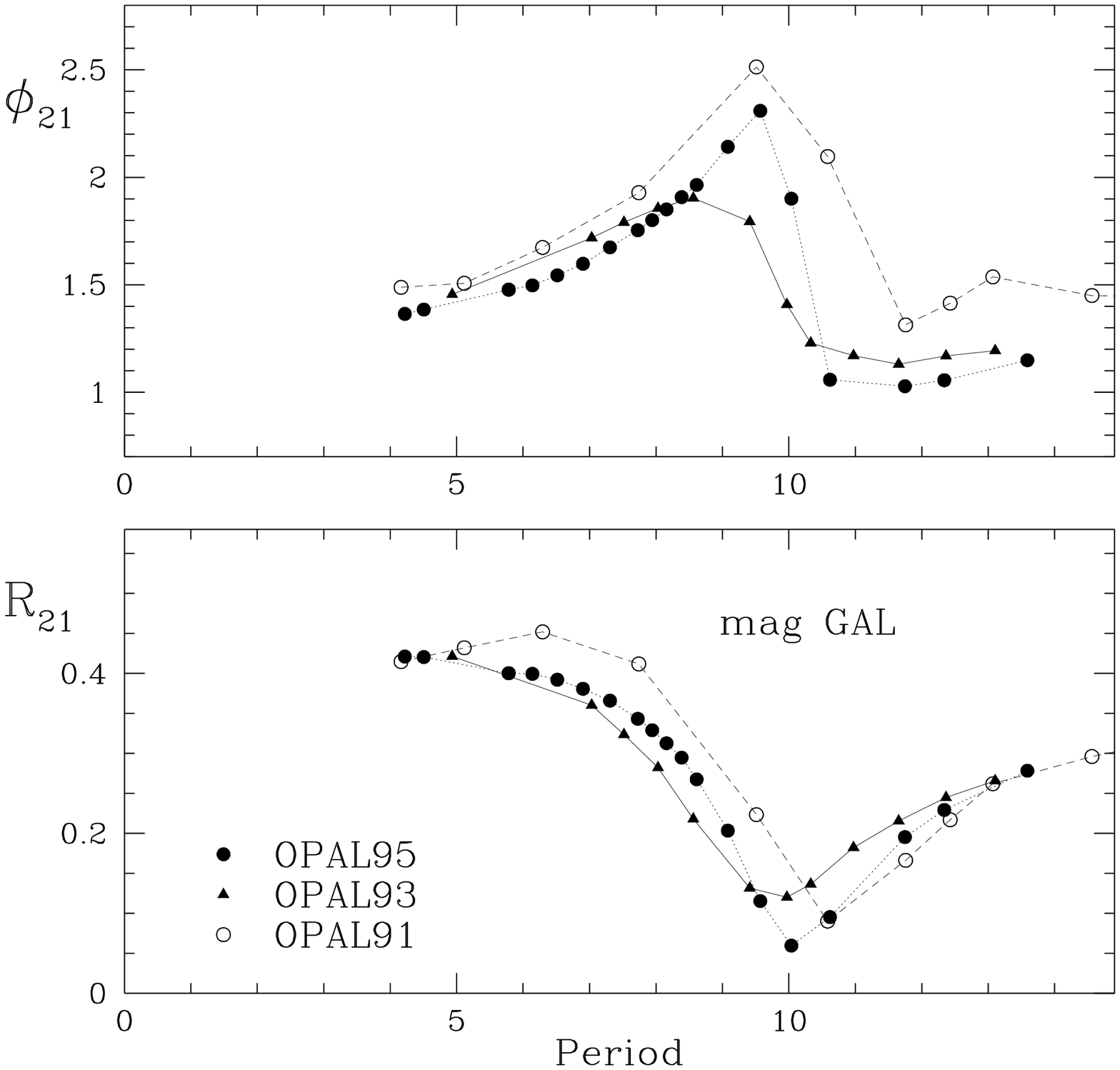,height=8cm}}
 \vskip -10pt
 \ni{\footnotesize Fig.~2\ Fourier decomposition coefficients, {\sl left}:
Effect of metallicity; {\sl right}: Effect of opacity.}
 \vskip 10pt

The observational data for the Galaxy \cite{SimonMoffett} show a considerably
larger scatter above $\approx\th$8$^d$.  Some of this spread is due to the
finite width of the instability strip, and to a spread in metallicity, with the
latter clearly dominating.  Taken in connection with the results of Fig.~2 {\sl
left}, this could be interpreted as indicating that the Galaxy has a large
dispersion in metallicity.

\subsection{Missing Fundamental Cepheids?}

Fig.~3 addresses the stability of the fundamental {\sl limit cycles}.  Shown
here is the Floquet stability coefficient corresponding to a perturbation with
the first overtone for a sequence that runs 100K to the left of the instability
strip.  The figure clearly indicates that {\sl the fundamental limit cycle is
unstable} in the range from $\approx\th$8--10$^d$ for all these sequences with
$Z=0.02$.  This result is largely independent of the opacity, in fact the
models of \cite{MBM} (OPAL91) were also unstable.  Preliminary results indicate
that for sequences located further away from the blue edge the instability of
the limit cycle decreases, but only slightly.  These results suggest that there
should be no or few fundamental pulsators with $Z=0.02$ in the period range
$\approx\th$8--10$^d$.

What does this then mean?  It turns out that the first overtone limit cycle is
stable in this regime, so that these stars must instead be first overtone
pulsators with periods $\approx\th$5.6--7$^d$.  Histograms of {\sl all}
observed Galactic Cepheids and of Andromeda \cite{BIT} indeed show a deficiency
of Cepheids short of the 10$^d$ period, and a corresponding increase short of
7$^d$ (This deficiency was left as an unexplained puzzle in ref. \cite{BIT}).
However, some fundamental Cepheids are observed in the 8--10$^d$ range.  These
stars should therefore have lower $Z$ values such as to make the fundamental
limit cycles stable.  (The corresponding histograms for the MCs do not display
these features, in agreement with their lower overall metallicity).  These
results corroborate our suggestion above that the Galactic Cepheids have a
sizeable metallicity dispersion.  An observational determination of the
metallicity as a function of period of both the fundamental and the first
overtone Galactic Cepheids would be an interesting test of these theoretical
predictions and a further check for the accuracy of the opacities.

\centerline{\psfig{figure=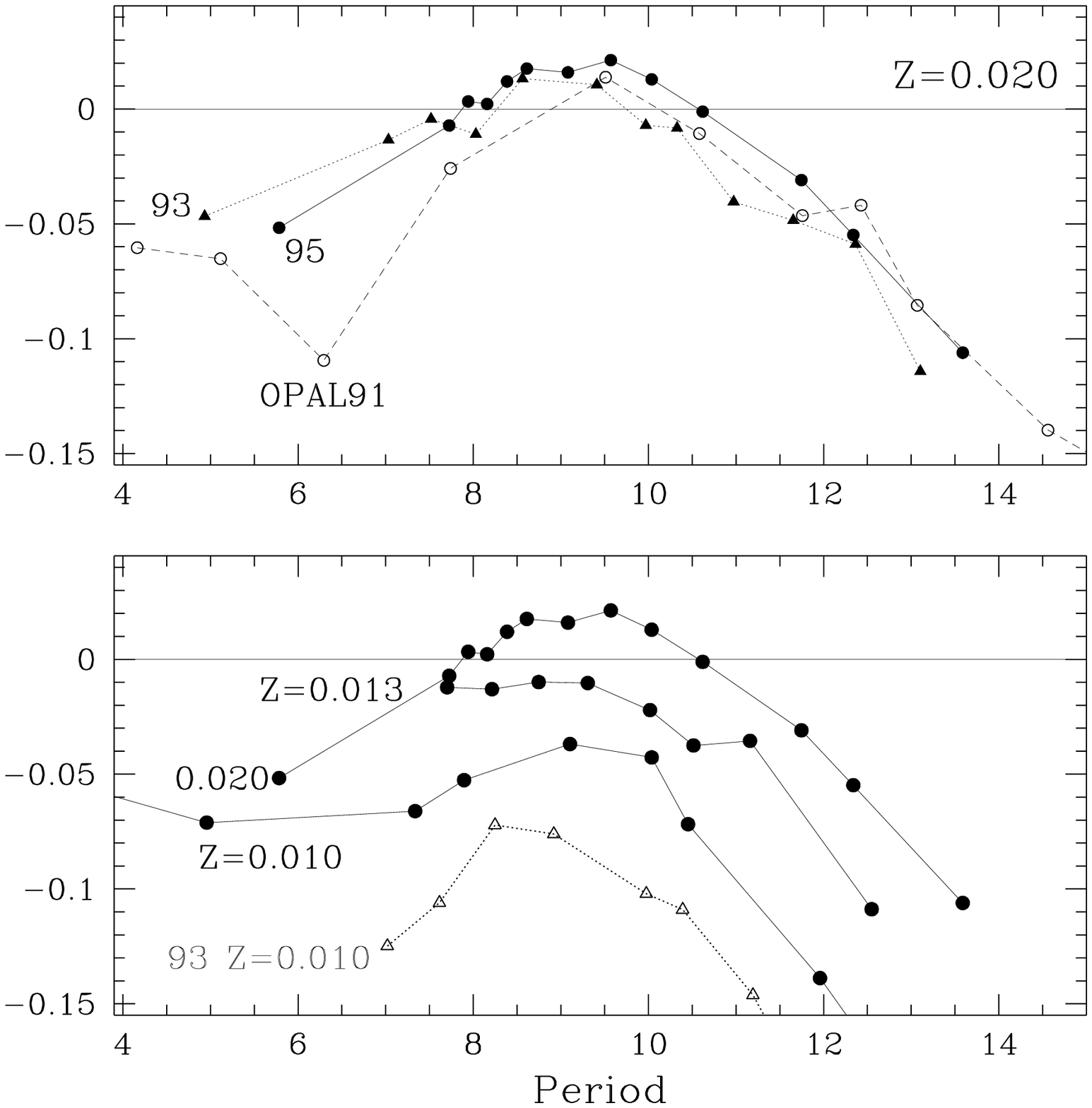,width=10cm}}

\ni{\footnotesize Fig.~3\ Linear stability (Floquet coefficient) of fundamental
limit cycle; {\sl top:} effect of opacity,\quad {\sl bottom:} effect of Z.}
 \vskip 10pt

\subsection{Uncertainties in the Opacities}

How sensitive are the nonlinear pulsations to the remaining uncertainties in
the opacities?  The right side of Fig.~2 gives a partial answer to this
question.  We show here the lowest Fourier parameters for three sequences of
models.  The two sequences, marked with triangles (OPAL93) and dots (OPAL95)
were both computed for a sequence whose $M$--$L$ relation was adjusted so that
the 2:1 resonance occurs at 10\dotd\ (with a slope of 1/3.56).  The open circle
sequence that has its resonance at $\approx\th$10\dotd 5\ represents sequence
$A$ of ref. \cite{MBM} which was computed with the first (1991) OPAL revision
of the opacities.  Our computations show that overall the effect of the change
from OPAL93 to OPAL95 is roughly equivalent to an increase of $Z$ from 0.02 to
0.03 (The reason is that the sensitivity is mostly due to the iron peak).  This
trend is also consistent with the results shown in Fig.~2.

\subsection{Overtone Cepheids}

The modelling situation is less comfortable for the first overtone
($s$~Cepheid) light curves.  The observational Fourier decomposition
coefficients display a very large excursion ($Z$ shape) in $\phi_{21}$ near
3$^d$.  This unambiguously indicates that a resonance is active in this region.
Is was conjectured by Antonello \& Poretti \cite{Antonello41} that the
structure is a result of a 2:1 resonance of the excited first overtone with the
fourth overtone.  However numerical hydrodynamical calculations \cite{AA}
\cite{SB} have failed to reproduce the observed features.  One of the problems
seems to be the fact that the fourth overtone is very strongly damped, and thus
hard to excite through nonlinear resonant coupling.  It would certainly help
theorists decipher the problem if first overtone radial velocity data could be
obtained and be Fourier decomposed.

\subsection{Beat Cepheids}

The so-called beat Cepheids are observed to pulsate in two modes
simultaneously, most of them in the fundamental and the first overtone (a few
also in the first and second overtones), with constant amplitudes and phases.
They are observed both in the Galaxy and in the MCs.  Hydrodynamics codes have
failed so far to produce beat behavior in the observed period range.  For this
behavior to occur both the fundamental and the first overtone limit cycles must
exist, but must be unstable ({\sl e.g.}  \cite{BuchlerMito}).  Numerical
simulations with radiative codes show that this destabilization does not occur
unless some resonance destabilizes the cycles.  The cause for beat behavior
remains a puzzle at the present time.

\section{RESONANCES} 

We have already discussed the dynamical effects of the bump resonance near
10$^d$.  Are there other resonances that play a dynamical role?  In order to
examine the presence of resonances we have computed a set of radiative Cepheid
models with masses ranging from 2 to 8\Mo, each with 9 $M$--$L$ relations and
effective temperatures varying from 7400--4600K.  In Fig.~4 we display the
linear period ratios of the middle $M$--$L$ relation (${\rm Lg} L = 3.56 {\rm
Lg} M + 0.775$) with $Z$=0.02.  Only those models are displayed in the left
(right) graph for which the fundamental (first overtone) is linearly unstable.
The reason for only showing unstable models is that this is a necessary (but
not sufficient) condition for the existence of a fundamental (first overtone)
limit cycle.  Furthermore, for the top curves ($P_{10}$ ($P_{21}$), because of
their usefulness for beat Cepheids we have only included those models which, in
addition, are unstable in the first (second) overtone.

The 'bump' resonance ($P_{20}$=1/2) for the fundamental Cepheids clearly stands
out in the left side figure at 10$^d$, as does the resonance ($P_{40}$=1/3)
near 7$^d$ which also plays a dynamical role, albeit much less important
\cite{BuchlerMito}.  Finally we note that there are further resonances, 4:1 and
5:1 at lower period, but with much higher overtones.  These modes could play a
dynamical role because the linear stability is not a monotonically increasing
function of mode number \cite{Glasner}.

  At the right edge of the figure is the half-integer resonance $P_{10}$=2/3
which gives rise to pulsations that are periodic, but repeat only every other
cycle (period-two behavior) \cite{MBreson}.  A recent analysis of Antonello
suggests that the star CC~Lyr might well exhibit this behavior.  The data are
very limited and additional observations of this star would be extremely
useful, especially if they could confirm the theoretical prediction
\cite{MoskBalter} that some Cepheids in the 20--25\dotd\ period range might
undergo regularly alternating cycles.

The Galactic beat Cepheids have a narrow range of period ratios, decreasing
from $P_{10}$=0.7097 for TU Cas ($P_0$=2\dotd 1393) to 0.6967 for V367 Sct
($P_0$=6.2929).  Both the range and the trend are consistent with the results
of Fig.~4.

In the right hand graph we show the period ratios relevant for first overtone
pulsators.  Again one sees immediately that the observed period ratio
$P_{21}$=0.8008 for the star CO~Aur at $P_1$=1.7830$^d$ is in good agreement
with the linear period ratios.  Of particular interest is the resonance
$P_{41}$=1/2 near 3$^d$ that was already mentioned in connection with the
$s$~Cepheids \cite{Antonello41}.  As for the fundamentals there are also
further, higher order resonances at lower period.  Again these resonances can
solely play a dynamical role if the corresponding overtone is only slightly
damped, which seems to occur around the 9th overtone\cite{Glasner}.

For lack of space we have only shown the period ratios for one $M$--$L$
relation.  It is obvious that these ratios display some sensitivity to
$M$--$L$.  Roughly speaking, a higher $M$--$L$ relation makes the period ratios
drop more rapidly with $P_0$.

\centerline{\psfig{figure=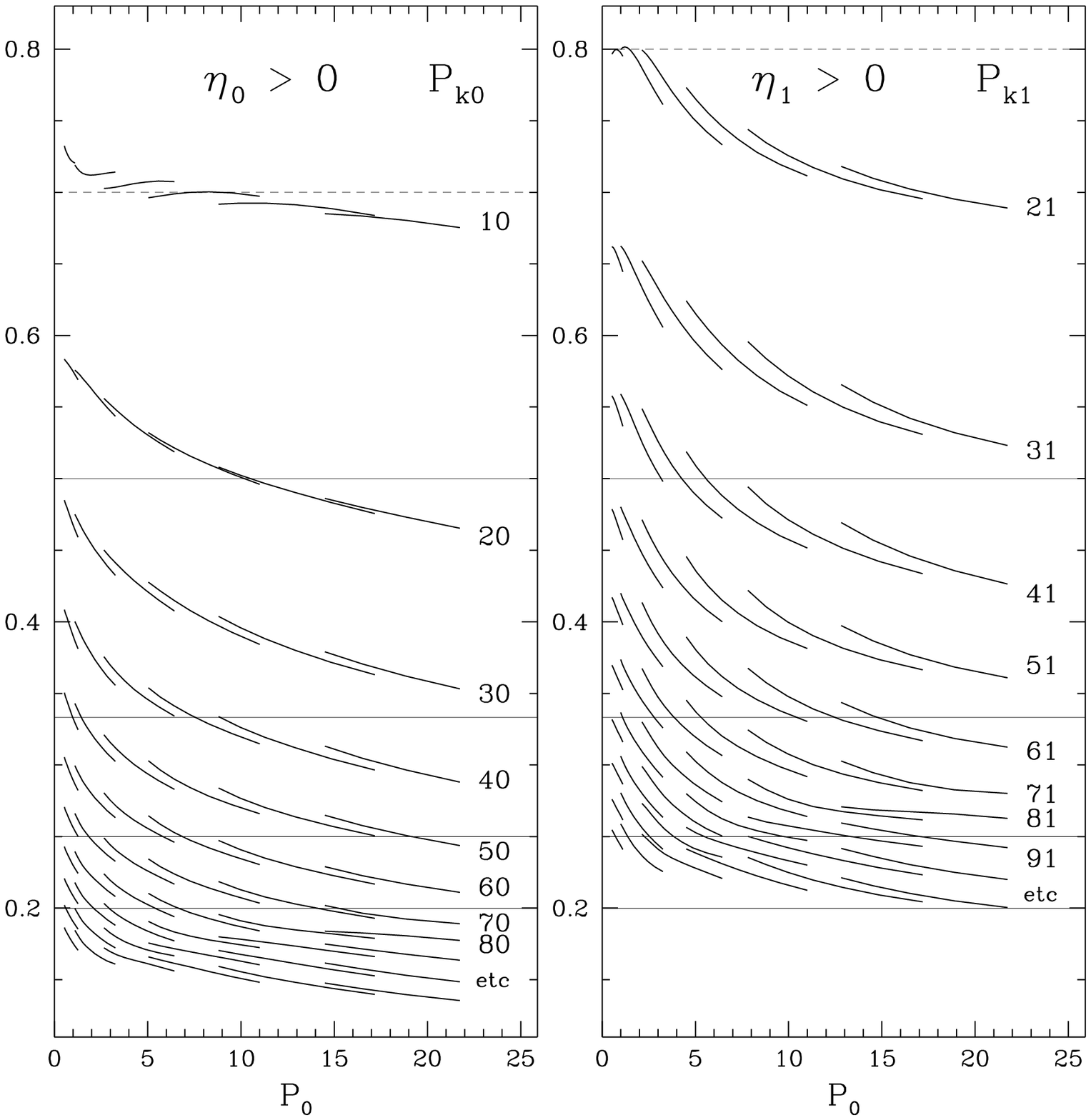,width=14cm}}

 \vskip -10pt

\ni{\footnotesize Fig.~4\ Linear period ratios, top to bottom $P_{10}$ to
$P_{11\th 0}$ and $P_{21}$ to $P_{11\th 1}$ for $X$=0.7, $Z$=0.02.
 }

 \vskip 20pt

 We have only considered two mode resonances here.  It does not appear at
present that multi mode resonances play a role in Cepheids (see however
\cite{KBlyrae} for RR ~Lyrae).

\subsection{Adaptive Codes}

Lagrangean codes have always been hampered by the fact that the hydrogen
ionization front is extremely narrow and that it undergoes substantial motion
through the star during the pulsation.  The resultant poor spatial resolution
has given rise to inaccuracies, as well as to bad jitter, especially in the
light curves.  In the last few years a number of groups have made use of modern
numerical adaptive mesh techniques \cite{Gehmyer} \cite{Dorfi} \cite{BKM} to
provide a good spatial resolution throughout the pulsation cycle.  In addition
these codes incorporate an option to solve the radiation hydrodynamics
equations instead of the usual radiation equilibrium diffusion equation.

The new codes provide much smoother and more accurate light curves.  However,
for the Cepheids and RR Lyrae, at least, the global properties, such as the
Fourier decomposition coefficients fortunately are not all that different.  The
same cannot be said for the Pop. II Cepheids where these numerical improvements
are essential.

\subsection{Convection}

Christy early on estimated that convection should be inefficient and the
convective flux should be small in the broad vicinity of the blue edge.  In
contrast, near the read edge convection must be essential and in fact it
determines the location of the red edge.  Convection has generally been ignored
in Cepheid models.  Recently Stellingwerf and Bono have advocated and used a
turbulence/convection recipe that seems to give quite reasonable results ({\sl
cf.} Bono in this Volume).

\subsection{Strange Modes}

A very careful look at Fig.~4 shows some irregularities in the behavior of the
period ratios for the higher overtones.  A thorough look has indicated that
this is not a numerical problem, but that it has a physical origin.  In Fig.~5
we show a blow-up for a typical example of a $M$=5\Mo, $L$=4348\Lo, $Z$=0.02
sequence (with OPAL95).  Interestingly and somewhat surprisingly, the figure
displays a level crossing that is quite familiar from 'strange' (radial) modes
\cite{SaioWheelCox} in luminous models, and from nonradial modes when $g$ and
$p$ modes overlap \cite{Unno}.  Here, it is due to the appearance of a strange
mode.  One can show that {\sl this mode has no direct counterpart in the
nonadiabatic vibrational spectrum}.  Interestingly this strange mode is
generally either linearly unstable or almost unstable.  Such a behavior was
uncovered some time ago \cite{Glasner}, but the strange nature of the mode was
not realized at the time.  A detailed study of these strange modes in Cepheid
models will be made in \cite{Buchlerstrange}.

 \vskip -2cm
 \centerline{\psfig{figure=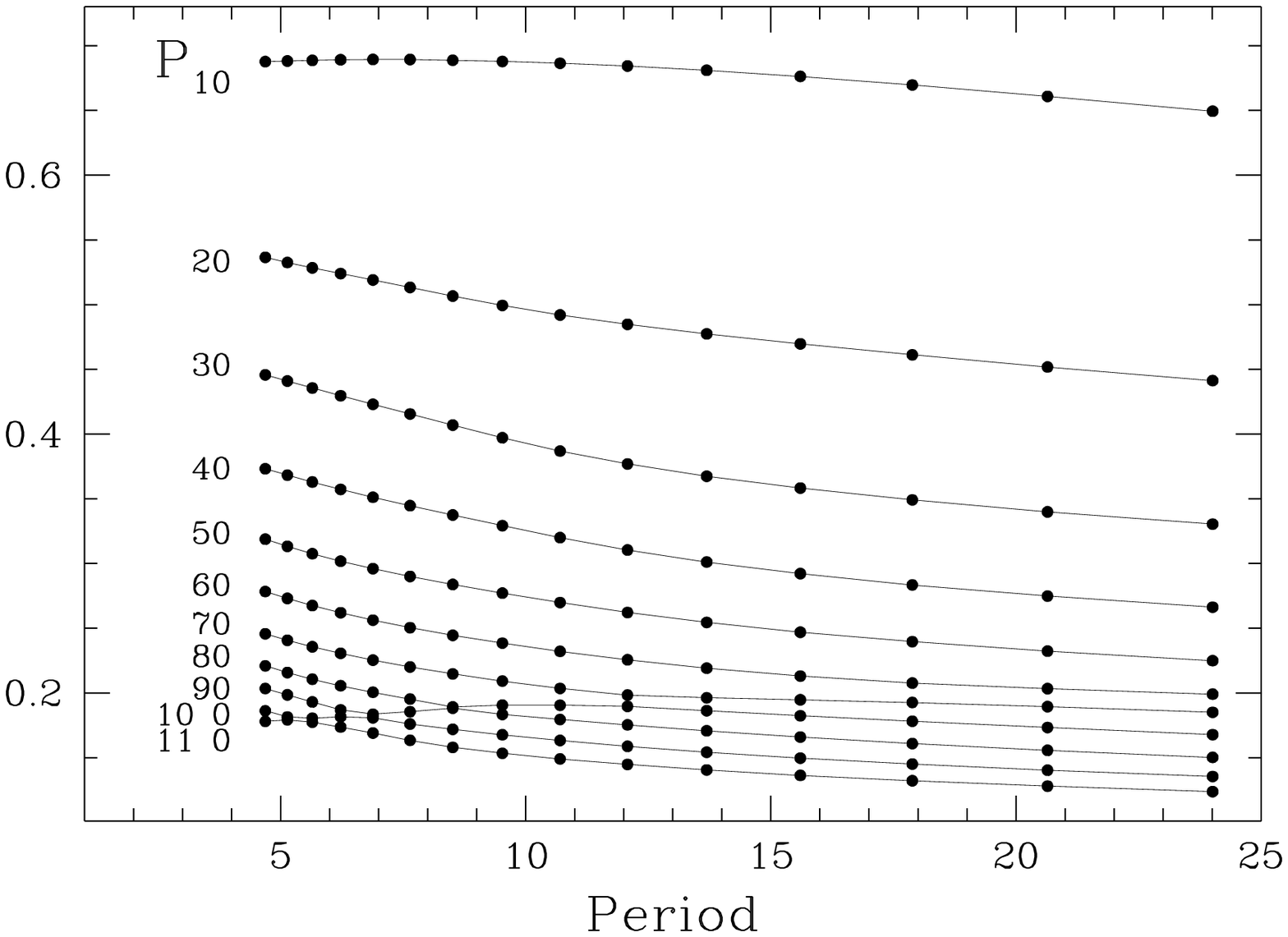,width=10cm}}

\ni{\footnotesize Fig.~5\ Period ratios $P_{k0}=P_k/P_0$ {\sl vs} $P_0$ for a
$M$=6\Mo, $L$=4348\Lo model, exhibiting the presence of a strange mode.}
 \vskip 10pt

Does the presence of this strange mode have any dynamical and observable
consequences?  Figs.~5 and 4 show that the strange mode can be in a 5:1
resonance with the fundamental mode, or in a 4:1 resonance with the first
overtone, perhaps simultaneously.  Because of its marginal stability and
excitability it could therefore play a role in destabilizing the limit cycle.
How resonances can destabilize the limit cycle is nicely illustrated in Fig.~1
of \cite{MBreson}.  This is clearly of interest in connection with the so far
unexplained beat behavior in Cepheids.  We have undertaken a systematical
numerical hydro survey to explore this conjecture.

\section{MAGELLANIC CLOUD CEPHEIDS}

We have pointed out above that the new opacities have seemingly reconciled
stellar evolution and pulsation with the observations \cite{MBM} \cite{KS} for
the {\sl Galactic} Cepheids.  However the new MC observations have recreated
havoc.

The EROS and MACHO projects have provided a plethora of Cepheid data in the
Magellanic Clouds (see this Volume).  The fact that the MC are metal
deficient by factors of 2 to 4 compared to the Galaxy makes these results
particularly interesting for understanding Cepheids.

In order to show the global effects of a reduced metallicity we have also
computed the linear properties of the same model sequences as in Fig.~4 but
with $Z$=0.01, more appropriate for the average metallicity of the LMC.  The
decrease in $Z$ causes an appreciable upward shift of all the $P_{k0}$, but a
much smaller shift in the $P_{k1}$.  The reason for this difference in shift is
that the metal opacities affect mostly the structure at high temperatures
($>$100\th 000K) where only the fundamental mode has a sufficiently large
amplitude.  The general trends of the shifts are in agreement with
observations, for example, the beat Cepheids have higher period ratios in the
MCs ({\sl cf.} Welch in this Volume; see also \cite{CD})).  Unfortunately,
however, the beat Cepheid data do not put any serious constraints on the
$M$--$L$ relations and on the models \cite{BKBG}.

A systematic and specific comparison for the bump Cepheids brings out
a severe problem \cite{BKBG}.  Most damning is the fact that the bump resonance
appears in approximately the same period range 9--11$^d$ and at approximately
the same luminosity in the LMC and the SMC.  This implies very low masses,
especially in the metal deficient SMC.  It also requires $M$--$L$ relations
that do not seem reconcilable with stellar evolution calculations (and the
current OPAL opacities).

\section{CONCLUSION}

The dark matter searches have provided us with wonderful new data on thousands
of variable stars in galaxies which are endowed by different metallicities.
The good news is that these data provide us with an enormously broadened data
base, but the bad news is that theorists have been temporarily sent back to the
drawing board.

\acknowledgements{
 It is a great pleasure to thank my current and past collaborators on Cepheids
for many exciting and fruitful discussions, in particular Marie-Jo Goupil,
Zoltan Koll\'ath, G\'eza Kov\'acs, Pawel Moskalik and Jean-Philippe Beaulieu.
I also thank Rick Piciullo for computing a few sequences of models on short
notice.  This work has been supported by NSF.
 }

\vfill
\end{document}